\documentclass{WileyMSP-template}
\usepackage{lipsum}
\usepackage{amsmath} 
\usepackage{subcaption}
\usepackage{textcomp}
\usepackage{comment}
\usepackage{xcolor}
\usepackage{multirow}   
\usepackage{makecell}
\usepackage{booktabs}
\usepackage{colortbl}
\usepackage{array}
\usepackage{rotating}

\begin{document}

\pagestyle{fancy}
\rhead{\includegraphics[width=1.2cm]{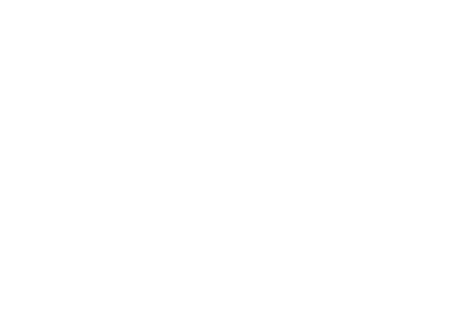}}

\title{High-Beam-Quality Meta-Grating Couplers for Large Collimated Free-Space Beams on Silicon-on-Insulator}

\maketitle

% Author: Please give full first and last names for authors and include * after the name of all corresponding authors

\author{Max Schittenhelm$^{1,2}$}
\author{Sebastian Häfner$^{1,2}$}
\author{Steffen Sauer$^{1,2,3}$}
\author{Stefanie Kroker$^{1,2,3,^*}$}

% Affiliations:
\begin{affiliations}
$^1$Institut für Halbleitertechnik, Technische Universität Braunschweig, Hans-Sommer-Str. 66, 38106 Braunschweig, Germany\\
$^2$Laboratory for Emerging Nanometrology, Langer Kamp 6a-b, 38106 Braunschweig, Germany\\
$^3$Physikalisch-Technische Bundesanstalt, 
Bundesallee 100, 
38116 Braunschweig, Germany\\
Email Address: $^*$s.kroker@tu-braunschweig.de
\end{affiliations}

% Keywords

% Abstract 

\begin{abstract}
Photonic integrated circuits on the silicon-on-insulator (SOI) platform typically interface with free space via grating couplers, but scaling these to collimated beams with diameters beyond 100\,\textmu m requires a fundamentally different regime of extremely weak, spatially distributed coupling. While such large-area couplers have been demonstrated, their beam quality has remained largely uncharacterized, even though applications such as coupling into high-finesse resonators or trapping of cold atoms require both a large aperture and a near-Gaussian profile. This article presents an SOI meta-grating coupler that emits collimated, near-Gaussian beams of approximately 300\,\textmu m waist diameter. The design synthesizes the required emission profile from a spatially tailored coupling strength, realized by locally varying a sub-wavelength unit cell while independently setting the local emission angle. This approach achieves the very low coupling strengths required for large beams and yields a measured beam quality of $M^2 \leq 1.10$. The scheme extends directly to other target profiles, such as flat-top or higher-order modes, rendering meta-grating couplers a practical chip-to-free-space interface for mode-matching-sensitive applications.

\end{abstract}

%%%%%%%%%%%%%%%%%%%%%%%%%%%%%%%%%%%%%%%
%%%%%%%% Introduction section %%%%%%%%%
%%%%%%%%%%%%%%%%%%%%%%%%%%%%%%%%%%%%%%%

\section{Introduction}

Photonic integrated circuits (PIC) based on the silicon-on-insulator (SOI) platform have become a cornerstone of modern integrated photonics. They combine compactness and mechanical robustness with access to the highly mature, CMOS-compatible fabrication technology \cite{Chen2018,Siew2021,Shekhar2024}. This combination has made SOI the platform of choice for a wide range of applications from optical communications \cite{Thomson2016} to sensing \cite{Vos2007, Milvich2021}, LIDAR \cite{Zhang2022, Wang2025} and quantum photonics \cite{Wang2020}.\\
A key enabling element for PICs is the interface between the on-chip and off-chip optical modes. Grating couplers (GCs) that connect the on-chip waveguides to off-chip optical fibers are now a well-established solution for this task and have been studied thoroughly \cite{Tamir1977, Taillaert2004, VanLaere2007, Su2018, Cheng2020, Zhong2025}. These grating couplers emit beams with diameters on the order of 10\,\textmu m to match the fiber's mode field diameter. Extending these couplers toward much larger beam radii and therefore toward collimated free-space beams, however, requires a fundamentally different coupling regime. The grating couplers must exhibit extremely small coupling strengths in order to extend the emission over several hundreds of microns.\\
Large beam size grating couplers have already found use in several applications such as 3D magneto-optical traps \cite{Luo2026, Isichenko2023, RoppLight2023}, sub-Doppler atomic spectroscopy \cite{Yulaev2024} or on-chip laser stabilization \cite{Hummon2018}. In all these cases, the beam merely has to illuminate an atomic ensemble or vapor cell over a large area, so that the detailed beam quality is of secondary importance and was neither systematically optimized nor analyzed. A second class of applications places considerably stricter demands on the emitted beam. Efficient coupling into optical resonators, in particular, relies on a high beam quality to achieve good mode matching \cite{Anderson1984}. Beyond that, applications involving trapped atoms, for instance the coherent addressing of individual sites in optical lattices, can equally benefit from large, well-defined beams with low divergence \cite{Blumenthal2024, Kosch2022, Tao2024}.\\ 
The beam quality relevant to these applications is commonly quantified by the beam quality parameter $M^2$, which describes how closely a beam approaches the diffraction-limited Gaussian case. A value of $M^2 = 1$ corresponds to an ideal Gaussian beam and thus sets the upper bound on the achievable overlap with a target mode such as that of an atomic trap or an optical resonator. While a low $M^2$ does not by itself guarantee good mode matching, which also depends on waist size, position, and wavefront curvature, it is a necessary precondition: no beam can exceed the overlap permitted by its $M^2$. It is therefore a practically meaningful figure of merit for evaluating large-area, chip-emitted beams.\\
Several groups have reported grating couplers achieving large beam diameters exceeding 100\,\textmu m \cite{ Kim2018, Ropp2021,  Yulaev2022, Livneh2022, Isichenko2023, Huang2023, Yulaev2023, BaronaRuiz2025}. To the best of our knowledge, however, none of these has combined such a large output diameter with a beam quality sufficient to meet the demands of the most mode-matching-sensitive applications.\\
In this work, we present and experimentally validate an SOI meta-grating coupler design scheme that achieves large, collimated beams with excellent beam quality factors of $M^2 \leq 1.10$. Two designs with emission angles of $-2$° and $-9$° are implemented, which require only a single etch step with minimum feature sizes of 229\,nm and 110\,nm, respectively. This performance is enabled by extending the sub-\linebreak wavelength grating coupler approach of [23] through a locally tailored coupling-strength profile, supported by an evanescent mode converter as demonstrated in [17].

\begin{figure}
  \centering  
  \includegraphics[width=0.8\linewidth]{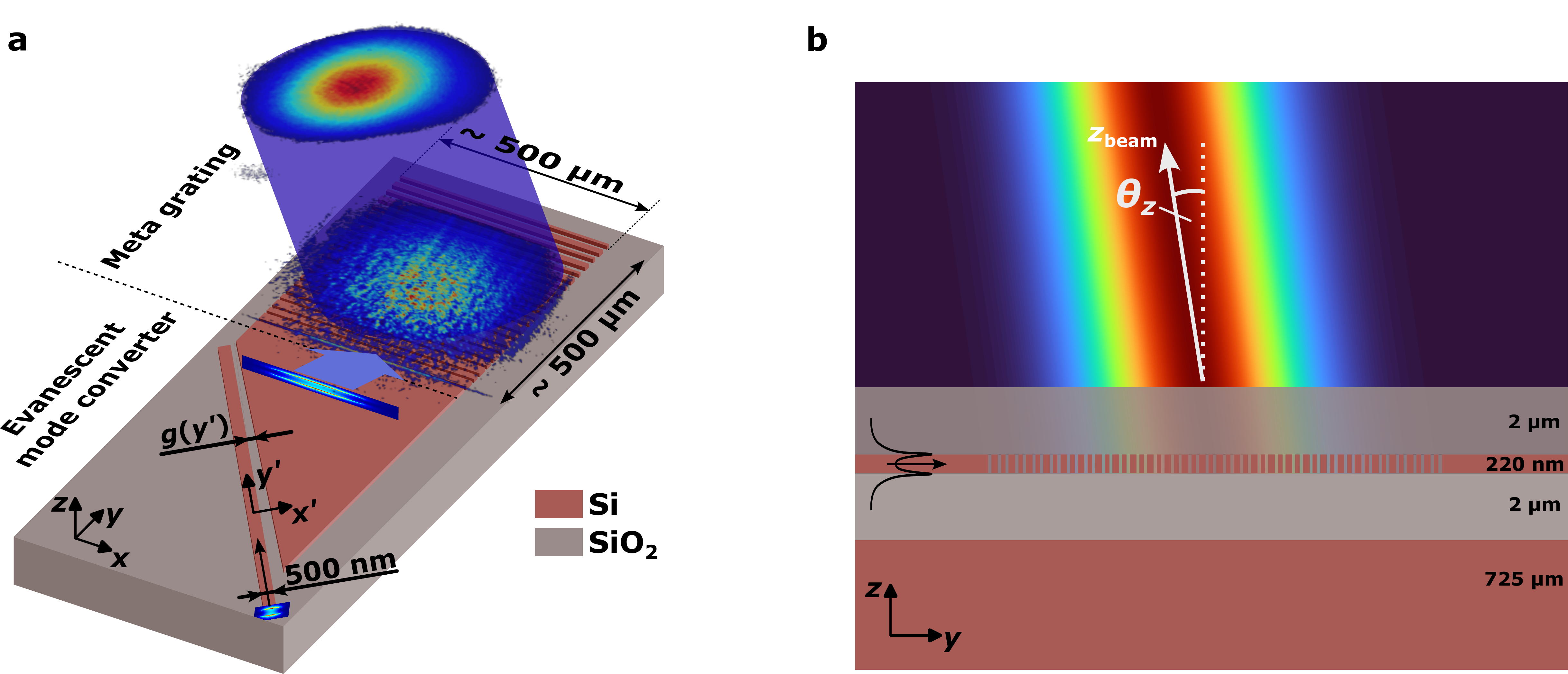}
  \caption{a) Schematic overview of the design consisting of an evanescent mode converter and a meta-grating coupler and b) the cross section of the grating coupler with the desired Gaussian beam emission. The portion of the beam that is emitted into the $-z$-direction is omitted for the sake of visual clarity.}
  \label{fig:grating_schematic}
\end{figure}

%%%%%%%%%%%%%%%%%%%%%%%%%%%%%%%%%%%%%%%
%%%%%%%%%%% Design section %%%%%%%%%%%%
%%%%%%%%%%%%%%%%%%%%%%%%%%%%%%%%%%%%%%%

\section{Design}

The presented designs consist of two parts: the evanescent mode converter which expands the tightly  confined mode from the waveguide (width of 500\,nm) to the slab (width of 500\,\textmu m) and the  meta-grating with variable coupling strength, as shown in Figure \ref{fig:grating_schematic}a. Together these two stages expand the mode field diameter to 300 \textmu m in the lateral and longitudinal direction, yielding a circular Gaussian beam. The TM00 mode was chosen as the operating mode since it is less confined along the z-direction compared to the TE00 mode, which allows for weaker interaction with the grating ridges and thus yields weaker coupling strengths. All designs discussed in this contribution use an SOI platform with a device layer thickness of 220 nm, as shown in Figure \ref{fig:grating_schematic}b, with an operating vacuum wavelength of $\lambda_0 = 1.55$ \textmu m.

%%%%%%%%%%%%%%
%%%%%%%%%%%%%% Sub-sec: Cherry-picked meta grating
%%%%%%%%%%%%%%

\subsection {Meta-grating coupler}\label{sec:meta-grating}
\begin{figure}
\centering  
\includegraphics[width=1\linewidth]{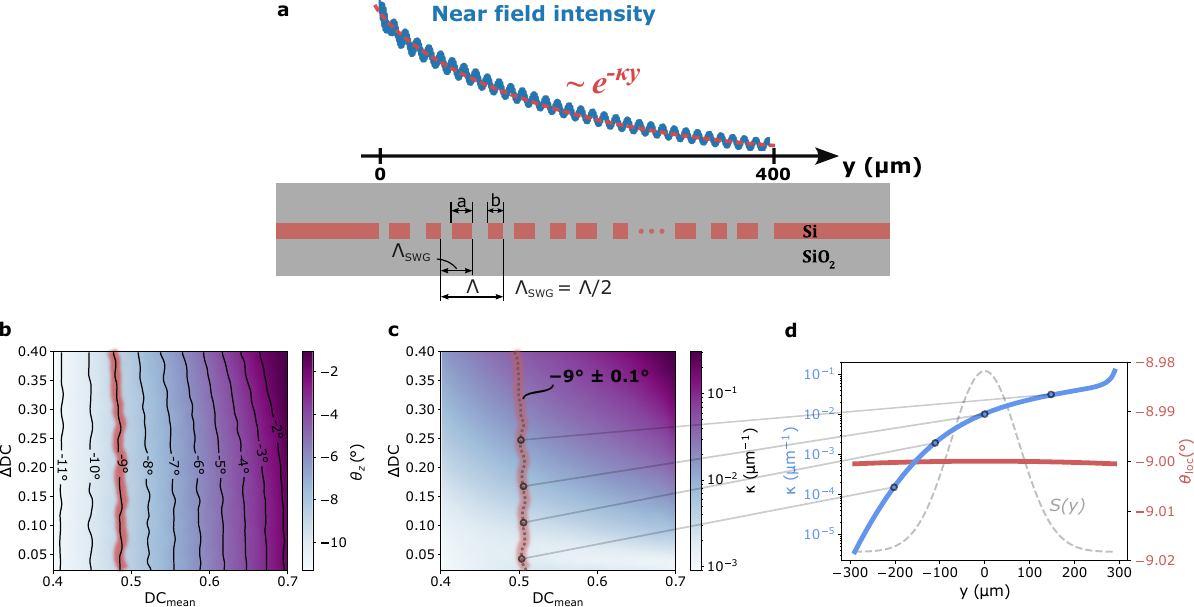}
\caption{a) Cross section of a homogeneous sub-wavelength grating with ridge widths a and b, which emits an exponentially decaying near field intensity profile. All  combinations of DC$_\text{mean} = \frac{a+b}{2\Lambda_\text{SWG}}$ and $\Delta$DC$=\frac{a-b}{\Lambda_\text{SWG}}$ are simulated and the emission angles $\theta_\text{z}$ and coupling strengths $\kappa$ are recorded in b) and c), respectively. A lookup-based inverse design method is employed by analytically calculating the required coupling strengths $\kappa(y)$ (blue line in d)) and local emission angles $\theta_\text{loc}(y)$ (red line in d)) along the grating and replicating them using the design sweeps from b) and c).}
\label{fig:cherry_picking}
\end{figure}

Our design scheme builds upon the sub-wavelength grating concept introduced by Barona-Ruiz et al. \cite{BaronaRuiz2025}, in which the coupling strength $\kappa$ of a grating coupler is strongly reduced by exploiting symmetry breaking in a sub-wavelength waveguide. For symmetric ridge widths (a = b) and a sub-wavelength period $\Lambda_\text{SWG} < \lambda_\text{mode}$, the structure acts as an effective-medium waveguide and does not radiate. Introducing a small asymmetry (a $\neq$ b) establishes a super-period $\Lambda = 2\Lambda_\text{SWG} > \lambda_\text{mode}$ that lies in the radiating regime, enabling emission profiles extending over several hundred micrometers \cite{BaronaRuiz2025}.
In a homogeneous grating, however, the unit cell, and therefore $\kappa$, remains constant along the propagation direction, so the emitted near field decays exponentially (Figure \ref{fig:cherry_picking}a). This fixes the emission profile and precludes the high beam quality required by mode-matching-sensitive applications. Here we lift this constraint: by spatially varying the unit cell, we realize an arbitrary, position-dependent coupling strength $\kappa (y)$ while independently controlling the local emission angle $\theta_\text{loc}(y)$. Combining the analytic $\kappa(y)$-profile synthesis established for conventional grating couplers \cite{Taillaert2004,Mekis2011,Mehta2017, Zhao2020} with the sub-wavelength approach of \cite{BaronaRuiz2025} allows us to reach the very low coupling strengths needed for large beams and, for the first time, to shape them into high-quality collimated Gaussian beams.
%%%%%%%% Figure start %%%%%%%%%
%%%%%%%% Figure end %%%%%%%%%

In our proposed design scheme, we aim to precisely replicate the local coupling strength $\kappa(y)$ required to emit a Gaussian beam. As has been shown in previous publications \cite{Taillaert2004,Mekis2011,Mehta2017, Zhao2020}, the local coupling strength required for a target emission profile $S(y)$ is given by
\begin{equation}
\kappa(y) = \frac{S(y)}{P(-\infty) - \int_{-\infty}^{y}S(t)dt}
\label{eq:kappa}
\end{equation}
where $P(-\infty)$ represents the initial power injected from the waveguide into the grating, with the boundary condition that all power is to be emitted into the beam:
\begin{equation}
\int_{-\infty}^{+\infty}S(y)dy \overset{!}{=} P(-\infty)  .  
\end{equation} 
2D simulations using the Lumerical FDTD: 3D Electromagnetic Simulator \cite{Lumerical_FDTD} were performed for various combinations of ridge widths $a$ and $b$ by sweeping the average duty cycle $\text{DC}_\text{mean} = (a+b)/2\Lambda_\text{SWG}$ and the absolute difference $\Delta \text{DC} = (a-b)/\Lambda_\text{SWG}$ over the specified regions in Figures \ref{fig:cherry_picking}b and \ref{fig:cherry_picking}c. Each point in these graphs corresponds to a homogeneous grating consisting of a repeating unit cell. For each single simulation the complex electrical near field  $E(y)$ was recorded above the grating. The near field intensities $|E(y)|^2$ were then fitted with an exponential function $\propto \text{exp}(-\kappa y)$ (see Figure \ref{fig:cherry_picking}a), from which the coupling strengths $\kappa$ were obtained for all designs. By projecting the complex near field $E(y)$ into the far field, the emission angle relative to the z-axis $\theta_z$ (see Figure \ref{fig:grating_schematic}b for reference) was obtained by finding the far field peak. The sweep results for the emission angles and coupling strengths of a sweep with a fixed periodicity of $\Lambda = 880$ nm are displayed in Figures \ref{fig:cherry_picking}b and \ref{fig:cherry_picking}c, respectively.\\
The desired Gaussian beam's electric field at the grating's plane can be analytically described by $E(y, z = 0) = |E(y,z=0)|\cdot \text{exp}(-i\cdot \phi(y))$, which for a beam waist of $w_0$ located at $z = 0$ and a propagation angle of $\theta_\text{z}$ with respect to the z-axis (see Figure \ref{fig:grating_schematic}b for reference) yields an intensity of
\begin{equation}
    S(y) \propto|E(y, z=0)|^2 = \Biggl|E_0\frac{1} {\sqrt{1 + \Bigl(\frac{\lambda y \mathrm{sin}(\theta_\text{z})}{\pi w_0^2}\Bigr)^2}} \cdot \mathrm{exp}\Biggl\{- \frac{(y\mathrm{cos}(\theta_\text{z}))^2}{w_0^2 \cdot \Bigl(1+ \bigl(\frac{\lambda y \mathrm{sin}(\theta_\text{z})}{\pi w_0^2}\bigr)^2\Bigr)}\Biggr\} \Biggr|^2
\label{eq:E_field}
\end{equation}
and
\begin{equation}
    \phi(y) = - \Biggl[\frac{2\pi}{\lambda} y\mathrm{sin}(\theta_\text{z}) + \frac{2\pi}{\lambda} \frac{(y\mathrm{cos}(\theta))^2}{2y\mathrm{sin}(\theta_\text{z})\cdot \Bigl(1 + \bigl(\frac{\pi w_0^2}{\lambda y\mathrm{sin}(\theta_\text{z})} \bigr)^2 \Bigr)}  - \mathrm{arctan}\Bigl(\frac{\lambda y\mathrm{sin}(\theta_\text{z})}{\pi w_0^2}\Bigr) \Biggr],
    \label{eq:phase}
\end{equation}
where $\lambda$ is the wavelength in the medium. If a focusing beam is desired, the equations above can be easily adjusted to account for the location of the waist to be at z $>$ 0. $|E(y, z=0)|^2$ from Equation \ref{eq:E_field} is then substituted for $S(y)$ in Equation \ref{eq:kappa} to calculate the resulting coupling strength $\kappa(y)$. Additionally, the phase expression $\phi(y)$ from Equation \ref{eq:phase} determines the local y-component $k_y(y)$ of the beam's wave number $k_0 = \frac{2\pi}{\lambda}$, from which in turn the required local emission angle $\theta_\text{loc}(y)$ can be calculated: 
\begin{equation}
k_y(y) = \frac{\mathrm{d}\phi(y)}{\mathrm{d}y} \Rightarrow \theta_{\mathrm{loc}}(y) = \mathrm{arcsin}\Bigl(\frac{k_y(y)}{k_0}\Bigr)
\end{equation}

%%%%%%%% Figure start %%%%%%%%%
\begin{figure}
\centering  
\includegraphics[width=1\linewidth]{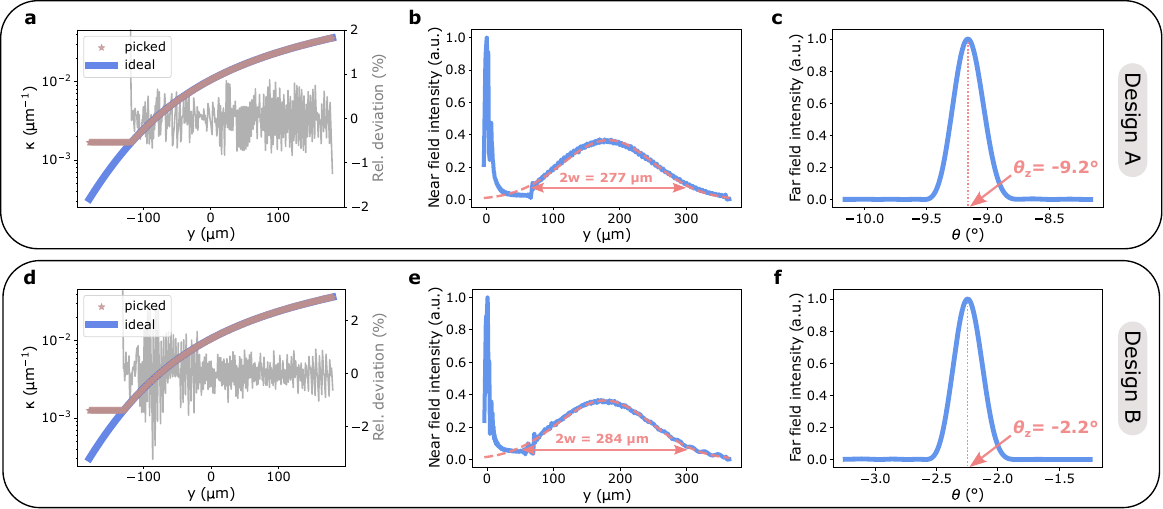}
\caption{Simulation data of both designs A and B with goal emission angles of $\theta_z^\text{A} = -9$° and $\theta_z^\text{B} = -2$°, respectively and a goal beam waist diameter of $2w_0$ = 300\,\textmu m. a) and d) custom-picked coupling strengths (orange stars) along the grating and the corresponding ideal values (blue line). b) and e) simulated near field intensity distributions predicting beam waist diameters of $2w_\text{0,A}$ = 277\,\textmu m and $2w_\text{0,B}$ = 284\,\textmu m. c) and f) corresponding far field projections, predicting emission angles of $\theta_{z\text{,sim}}^\text{A} = -9.2$° and $\theta_{z\text{,sim}}^\text{B} = -2.2$°. The simulated radiation efficiencies are $T_\text{sim}^\text{A} = T_\text{sim}^\text{B} = 70\,\%$.}
\textsuperscript{}
\label{fig:simulation_data}
\end{figure}
%%%%%%%% Figure end %%%%%%%%%
In this contribution, two designs A and B are evaluated. Both are designed to have a goal beam waist diameter of 2$w_0$ = 300 \textmu m. We choose goal emission angles of $\theta_z^\text{A} = -9$° and $\theta_z^\text{B} = -2$° to avoid the strong Bragg reflection associated with vertical coupling and to demonstrate the feasibility of creating custom emission angles. The periods for design A and B were chosen to be $\Lambda_\text{A} = 880$\,nm and  $\Lambda_\text{B} = 900$\,nm, respectively. Representative for both designs, the entire design process for design A is shown in Figures \ref{fig:cherry_picking}b-d. The resulting constant local emission angle of $-9$° in Figure \ref{fig:cherry_picking}d allows to filter out all homogeneous designs, i.e. combinations of ridge widths $a$ and $b$, that exhibit an emission angle of $-9\pm0.1$° in the coupling-strength map in Figure \ref{fig:cherry_picking}c. Subsequently, individual unit cells are placed along the y-direction to best fit the required $\kappa(y)$-function, as indicated by the gray lines in between Figures \ref{fig:cherry_picking}c and \ref{fig:cherry_picking}d.\\
Figures \ref{fig:simulation_data}a and \ref{fig:simulation_data}d show the analytically calculated coupling strength values $\kappa(y)$ (blue line) and the custom picked coupling strengths (orange stars) from the simulation sweeps (Figure \ref{fig:cherry_picking}c) for designs A and B, respectively. For small $\kappa$-values below $2\cdot10^{-3}$\,\textmu m$^{-1}$, the design spaces lacked sufficiently small $\kappa$-values, leading to a mismatch at the beginning of the gratings in both cases. However, for the remainder of the gratings, the coupling strengths could be neatly matched with small deviations. This directly translates to the simulated near field intensity profiles in  Figures \ref{fig:simulation_data}b and \ref{fig:simulation_data}e, which only deviate from a Gaussian shape at the beginning where the required coupling strengths did not match. The near field intensity peak at $y = 0$\,\textmu m is attributed to the jump in refractive index occurring at the slab-grating interface  and causes a parasitic, non-directional upwards emission of 13\,\% and 10\,\% for designs A and B, respectively, as shown in the supplementary materials. This can be mitigated by employing a gradual transition region, as shown in \cite{BaronaRuiz2025}. The simulated far field emission angles (Figures \ref{fig:simulation_data}c and \ref{fig:simulation_data}f) of both designs were determined to be  $\theta_\text{z,sim}^\text{A} = -9.2$° and $\theta_\text{z,sim}^\text{B} = -2.2$°, closely matching the goal angles of $\theta_\text{z}^\text{A} = -9$° and $\theta_\text{z}^\text{B} = -2$°, respectively, while showing radiation efficiencies of $T_\text{sim}^\text{A}$ =70\,\% and $T_\text{sim}^\text{B}$ = 70\,\%. Since the proposed design scheme uses fully etched gratings the emission from the grating is equally split into the positive and negative z-directions. The light emitted downwards experiences some reflection at the high-index-contrast interface of the buried oxide layer and the Si handle layer. Depending on the emission angle and thickness of the buried layer, the reflected light interferes with the light which is emitted directly upwards from the chip, leading to simulated radiation efficiencies upwards of 50\,\%. This value can still be optimized if the buried oxide layer thickness can be chosen arbitrarily. In our case, this value was fixed to 2 \textmu m due to the fabrication process used.

%%%%%%%%%%%%%%
%%%%%%%%%%%%%% Sub-sec: Evanescent mode converter
%%%%%%%%%%%%%%

\subsection{Evanescent mode converter}
%%%%%%%% Figure start %%%%%%%%%
\begin{figure}
\centering  
\includegraphics[width=1\linewidth]{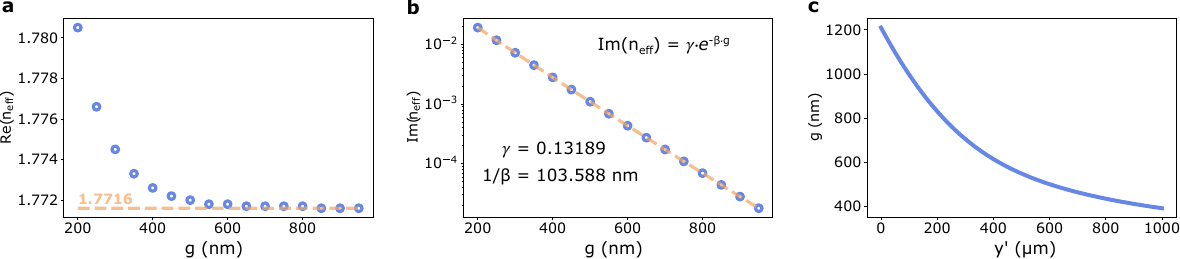}
\caption{Real (a) and imaginary (b) parts of the TM00 effective mode index as a function of the gap in between the waveguide and the slab region obtained through COMSOL Multiphysics\textsuperscript{\textregistered} FEM simulations \cite{COMSOL}. c) The analytically derived gap distance as a function of the propagation distance y' along the waveguide.}
\textsuperscript{}
\label{fig:evanescent_coupler}
\end{figure}
%%%%%%%% Figure end %%%%%%%%%

The longitudinal  beam shape is controlled by the grating design presented above. To achieve the same beam profile in the transverse direction (along the x-coordinate in Figure \ref{fig:grating_schematic}a), an evanescent mode converter was designed using the design approach from \cite{Kim2018}. For the sake of thoroughness, the key steps are explained here. The FEM solver from COMSOL Multiphysics\textsuperscript{\textregistered}' Wave Optics Module \cite{COMSOL} was used to calculate the waveguide's 2D mode profile as a function of the gap distance $g$ between the waveguide and the slab region. The waveguide itself was assumed to be lossless and the radiation of the mode's energy into the slab due to evanescent tunneling was accounted for by introducing optical losses in the slab. For each gap distance, the real and imaginary parts of the mode's effective index were recorded and are displayed in Figures \ref{fig:evanescent_coupler}a and \ref{fig:evanescent_coupler}b, respectively. The real part of the mode index will determine the emission angle $\theta_\text{slab}$ into slab according to 
\begin{equation}
    \text{cos}(\theta_{\text{slab}}) = \Re\{n_{\text{eff}}^{\text{wg}}\} /\Re\{n_{\text{eff}}^{\text{slab}}\}
\label{eq:slab_angle}
\end{equation}
and the imaginary part determines the power loss factor of the mode $\alpha(y')=\frac{4\pi}{\lambda_0}\Im\{n_{\text{eff}}^{\text{wg}}\}$ \cite{Kim2018}. 

The resulting gap function for a Gaussian slab mode $S(y') \sim \text{exp}(-2\frac{(y'-y'_0)^2}{\widetilde{w}^2})$ with waist $\widetilde{w}$ and center position $y'_0$ is then given by:

\begin{equation}
    g(y') = -\frac{1}{\beta}\text{ln} \Bigg( 
    \frac{\lambda_0}{\sqrt{2}\gamma\widetilde{w}\pi^{3/2}}
    \frac{e^{-2\frac{(y'-y'_0)^2}{\widetilde{w}^2}}}{1 - \text{erf}\Bigl\{ \sqrt{2} \frac{y'-y'_0}{\widetilde{w}}  \Bigr\} }
    \Biggr)
\end{equation}

with $\beta, \gamma$ being the fit coefficients from Figure \ref{fig:evanescent_coupler}b. For an effective slab beam waist diameter of $2w_\text{slab} = 300$\,\textmu m, the slab emission angle needs to be taken into account by setting the waist in the equation above to $\widetilde{w} = w_\text{slab}/\text{sin}(\theta_\text{slab})$. The real part of the mode index and therefore the slab emission angle was assumed to be constant, i.e. Equation \ref{eq:slab_angle} then yields $\theta_\text{slab} = \text{cos}^{-1}(1.7716 / 2.0537) \approx 30.36$°, which is a good approximation due to the resulting gap profile exhibiting values larger than 400\,nm (Figure \ref{fig:evanescent_coupler}c).
%%%%%%%%%%%%%%%%%%%%%%%%%%%%%%%%%%%%%%%
%%%%%%%% Experimental results %%%%%%%%%
%%%%%%%%%%%%%%%%%%%%%%%%%%%%%%%%%%%%%%%

\section{Experimental Results}
%%%%%%%% Figure start %%%%%%%%%
\begin{figure}
\centering  
\includegraphics[width=1\linewidth]{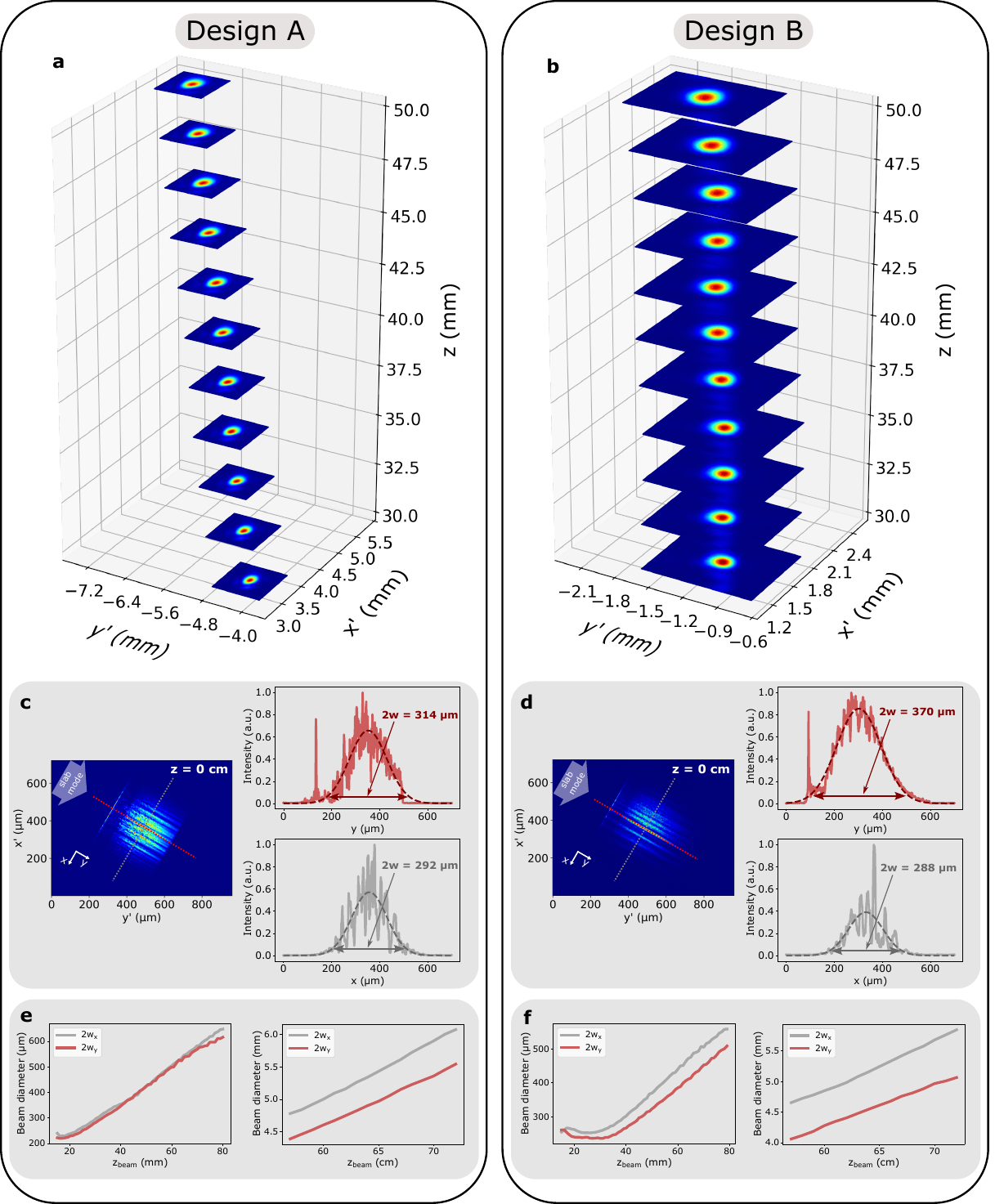}
\caption{Experimental beam analysis for both designs. a) and b) show the imaged beam intensity slices from 30 mm to 50 mm above the chips. c) and d) show the analysis of the near field intensity along (gray lines) and perpendicular (red lines) to the grating ridges. e) and f) show the beam waist diameters along x and y acquired from Gaussian fits of the line profiles  with maximum 2$\sigma$-uncertainties of $\pm$8.28 \textmu m  and $\pm$76.6 \textmu m for the plots with 15 mm $< \text{z}_\text{beam} <80$ mm and 57 cm $< \text{z}_\text{beam}<72$ cm, respectively.}
\textsuperscript{}
\label{fig:full_beams}
\end{figure}
%%%%%%%% Figure end %%%%%%%%%
To experimentally validate the design process described above, the two different grating designs A and B with design emission angles of $\theta_z^\text{A} = -9$° and  $\theta_z^\text{B} = -2$° and design waist diameters of $2w_0 = 300$ \textmu m were fabricated using the Applied Nanotools Inc. NanoSOI Fabrication Service \cite{AppliedNanotools}, as described in the Methods section. The evanescent mode converter was identical for both designs and exhibited the gap function from Figure \ref{fig:evanescent_coupler}c.\\
The beam cross sections were imaged from $z = 0$\,mm (near field) up to $z = 80$\,mm in 1\,mm steps. For visual clarity, the beam slices are shown from $z = 30$\,mm up to $z = 50$\,mm in 2\,mm steps in Figures \ref{fig:full_beams}a and \ref{fig:full_beams}b. By tracking the beam centers, the emission angles were determined to be $\theta_\text{z,exp}^\text{A} = -9.4$° and $\theta_\text{z,exp}^\text{B} = -2.7$°, closely matching the simulated emission angles $\theta_\text{z,sim}^\text{A} = -9.2$° and $\theta_\text{z,sim}^\text{B} = -2.2$°. 
The radiation efficiencies were determined to be $T_\text{exp}^\text{A}$ = 31\,\% and $T_\text{exp}^\text{B}$ = 49\,\% (see Methods section for details), below the simulated value of $T_\text{sim}$=70\,\%. 
Part of this reduction is attributed to the parasitic emission at the slab-grating interface. As shown in the supplementary materials, this non-directional emission is simulated to amount to 13\,\%
and 10\,\% for designs A and B, respectively, and can be mitigated by a gradual transition region like the one shown in \cite{BaronaRuiz2025}.
For fully-etched gratings, the emission is in principle split equally into the 
$+z$ and $-z$ directions, which would limit the upward radiation efficiency to 50\,\%. The reflection at the BOX-substrate interface can lift this limit: depending on the emission angle and the oxide thickness, the downward-emitted light interferes constructively with the upward emission and allows efficiencies above 50\,\%. The remaining gap between the measured efficiencies and this expectation is, however, dominated by the way the radiation efficiency is extracted rather than by the coupler itself. As detailed in the Methods section, the efficiency is obtained by taking into account the estimated fiber-array-to-chip coupling efficiency, which relies on vendor-specified parameters. (For design A, only TE-polarization measurement data was available. For TM-polarization, the vendor estimated an additional loss of 0.5 dB) Further, designs A and B were located on separate chips with different edge-coupling schemes, so their absolute efficiency values are not directly comparable, and the value for design B represents a lower bound. We therefore treat these figures as order-of-magnitude estimates carrying a substantial systematic uncertainty. Notably, the central result of this work, the beam quality, is measured directly from the free-space beam and is independent of these in-coupling estimates.\\
Figures \ref{fig:full_beams}c and \ref{fig:full_beams}d show the respective near field images of the beams with line profiles corresponding to the field distributions caused by the grating design (red profiles, along y) and the evanescent mode converter (gray profiles, along $x$). In both cases, Gaussian-like intensity distributions are observed for the red y-profiles, with waist diameters of 
314\,\textmu m and 
370\,\textmu m, respectively. While design A matches the target waist of 
300\,\textmu m to within 5\,\%, the larger deviation of design B is most likely related to its shallower emission angle of $-2$°, which lies closer to the vertical Bragg condition. In this regime the local coupling strength is more sensitive to fabrication-induced deviations, and the initial region of insufficient $\kappa$-matching (Section \ref{sec:meta-grating}) has a stronger effect on the resulting beam width, broadening the effective waist along the grating direction. The distinct emission peak at the beginning of the grating in the red profiles can be attributed to the jump in the effective mode index that occurs at the slab-grating interface and causes parasitic emission of the order of 10\,\% (see supplementary information). The transverse $x$-profiles, set by the evanescent mode converter, are unaffected by this and exhibit Gaussian-like distributions with waist diameters of  292\,\textmu m and 288\,\textmu m, respectively, closely matching the goal waist of 
300\,\textmu m for both designs. Additionally, an interference pattern along the $x$-profiles can be observed. This we attribute to the slab mode not being perfectly vertically incident to the grating coupler. This is likely caused by the evanescent mode converter design: a deviation of the simulated waveguide and slab mode indices from the real ones causes a change in the propagation angle of the slab mode according to Equation \ref{eq:slab_angle}. This causes a non-zero $x$-component of the slab wave vector along the grating ridges, which oscillates as a standing wave along the $x$-direction.
%%%%%%%% Figure start %%%%%%%%%
\begin{figure}
\centering  
\includegraphics[width=0.8\linewidth]{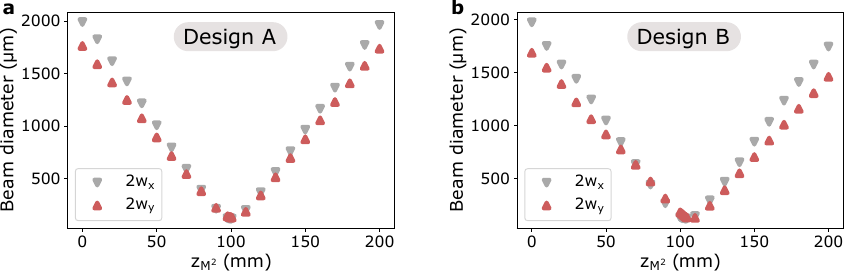}
\caption{ISO 11146 \cite{ISO11146-1} compliant $M^2$ measurements for a) design A and b) design B. The beams emitted from the chip were collimated using a plano-convex lens with a focal length of $f_\text{c} = 500$\,mm placed at a distance of $\approx$ 50\,cm from the chip and were then focused inside of the $M^2$ measurement system with a second plano-convex lens with $f_{\text{M}^2} = 250$\,mm. A spherical aperture with a diameter of 8\,mm was used to filter out side lobes which appeared along the x-directions. Measurements along the x- and y-axes correspond to the evanescent mode converter and the grating, respectively. The spatial coordinate $z_{\text{M}^2}$ refers to the position of the linear stage within the system. The ISO-compliant hyperbolic fits yield values of M$^2_\text{x,A} = 1.04$, M$^2_\text{y,A} = 1.10$, $M^2_\text{x,B} = 1.05$ $M^2_\text{y,B} = 1.09$ with an uncertainty of $\pm$5\%.}
\textsuperscript{}
\label{fig:M2}
\end{figure}
%%%%%%%% Figure end %%%%%%%%%
Since the aim is to demonstrate collimated Gaussian beams, the analysis of the far field emissions of both gratings are of interest. Figures \ref{fig:full_beams}e and \ref{fig:full_beams}f show the evolution of the beam waist diameters of the beams along both the x- and y-axes as a function of the beam propagation distance $z_\text{beam}$. For each image, the center position of the beam and the beam waists along the two grating axes $x$ and $y$ were determined by fitting a Gaussian $\propto \text{exp}({-2(n-n_0)^2/w_n^2}), n = x,y$ to the corresponding line profiles. Due to the non-vertical emission angle, the z-coordinate was adjusted to point along the propagation direction, $z_\text{beam} = z / \text{cos}(\theta_\text{z,exp})$, and the waist diameter along the x-axis was corrected accordingly, $2w_\text{x} = 2w_\text{x,fit}\cdot \text{cos}(\theta_\text{z,exp})$. The graphs in Figures \ref{fig:full_beams}e and \ref{fig:full_beams}f show the analysis for 15\,mm $< z <$ 80\,mm and 57\,cm $< z_\text{beam} <$ 72\,cm. Since the images up to a height of approximately 15 mm showed near field interference patterns, which prohibited proper Gaussian fits and beam center calculations, the analysis starts at a height of $z_\text{beam} = 15$\,mm.  The analysis of the beams up to a height of 80\,mm was conducted using the microscope setup with a 10X objective, where as for the analysis in the range of 57\,cm to 72\,cm the SWIR camera was directly placed in the beam path without any magnifying optics.\\
In order to quantify the quality of the beams generated by the two grating designs, ISO 11146 \cite{ISO11146-1} compliant $M^2$ measurements were conducted. The beam quality factor $M^2$ is defined by
\begin{equation}
    M^2 = \frac{2w_m\Theta_m}{2w_0\Theta_0} \stackrel{w_m=w_0}{=} \frac{\Theta_m}{\Theta_0}
\end{equation}
where $w_m$ and $\Theta_m$ represent the measured beam waist and divergence angle and $\Theta_0$ the beam waist and divergence angle of an ideal Gaussian beam. $M^2$ thus quantitatively describes how close the beam profile is to an ideal Gaussian. The measurements were conducted using a commercial $M^2$ Measurement System (see Methods section) which employed a scanning slit beam profiler. The beam profiler's measurement axes were aligned to the axes corresponding to the grating design's emission profile (y-axis in Figure \ref{fig:grating_schematic}a) and the evanescent mode converter's emission profile (x-axis in Figure \ref{fig:grating_schematic}a), respectively. Along the x-direction, when focusing the beam during the $M^2$-measurement, the beams' side lobes significantly interfere with the main profile around the focal position, prohibiting proper $M^2$ measurement of the x-axes. Therefore, a spherical aperture with a diameter of 8\,mm was placed in the beam path (see Figure \ref{fig:experimental_setup}c) to suppress these side lobes. The resulting beam quality factors in the x-direction (corresponding to the evanescent mode converter) and y-direction (corresponding to the grating design) of designs A and B were determined to be $M^2_\text{x,A} = 1.04$, $M^2_\text{y,A} = 1.10$, $M^2_\text{x,B} = 1.05$ $M^2_\text{y,B} = 1.09$ with an uncertainty of $\pm$5\,\%, thus showing excellent beam properties. Table \ref{tab:comparison} shows a comparison of the goal, simulated and experimental values of both designs, which overall yield excellent agreement. 
%%%%%%%%%%% BEGIN TABLE
\definecolor{rowgray}{gray}{0.93}
\definecolor{lightgreen}{rgb}{0.85, 0.95, 0.85}

\begin{table}[]
\centering
\renewcommand{\arraystretch}{1.3}
\caption{This table compares the simulation and experimental results of both designs regarding their emission angles $\theta_\text{z}$, near field waist diameters 2$w_0$, radiation efficiency $T$ and beam quality factors $M^2$ ($\pm 5\,\%$) along the x- and y-axes.}
\begin{tabular}{l l ccccc}
\toprule
& & $\theta_\text{z}$ (°) & $2w_0$ (\textmu m) & $T$ (\%) & \cellcolor{lightgreen}$M^2_x$ & \cellcolor{lightgreen}$M^2_y$ \\
\midrule
\multirow{3}{*}{\rotatebox[origin=c]{90}{\textbf{Design A}}}
    & Goal       & $-9$    & $300$ & $>50$     & \cellcolor{lightgreen}$1.0$ & \cellcolor{lightgreen}$1.0$ \\
    & \cellcolor{rowgray}Simulation & \cellcolor{rowgray}$-9.2$ & \cellcolor{rowgray}$277$ & \cellcolor{rowgray}$70$    & \cellcolor{lightgreen}$-$  & \cellcolor{lightgreen}$-$   \\
    & Experiment & $-9.4$  & $314$ & $31$      & \cellcolor{lightgreen}$1.04$ & \cellcolor{lightgreen}$1.10$ \\
\midrule[1pt]
\multirow{3}{*}{\rotatebox[origin=c]{90}{\textbf{Design B}}}
    & Goal       & $-2$    & $300$ & $>50$     & \cellcolor{lightgreen}$1.0$  & \cellcolor{lightgreen}$1.0$ \\
    & \cellcolor{rowgray}Simulation & \cellcolor{rowgray}$-2.2$ & \cellcolor{rowgray}$284$ & \cellcolor{rowgray}$70$    & \cellcolor{lightgreen}$-$  & \cellcolor{lightgreen}$-$   \\
    & Experiment & $-2.7$  & $370$ & $\geq 49$ & \cellcolor{lightgreen}$1.05$ & \cellcolor{lightgreen}$1.09$  \\
\bottomrule
\end{tabular}
\label{tab:comparison}
\end{table}
%%%%%%%%%%% END TABLE

%%%%%%%%%%%%%%%%%%%%%%%%%%%%%%%%%%%%%%%
%%%%%%%%%%%%% Methods %%%%%%%%%%%%%%%%%
%%%%%%%%%%%%%%%%%%%%%%%%%%%%%%%%%%%%%%%

\section{Methods}

\begin{figure}
\centering  
\includegraphics[width=0.8\linewidth]{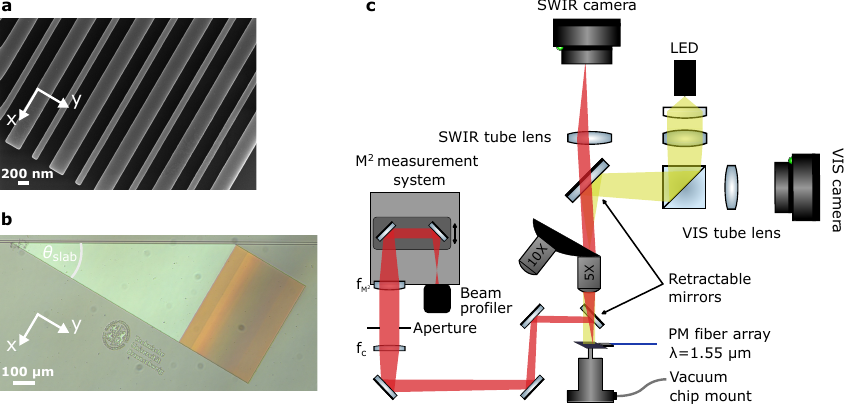}
\caption{a) Scanning electron microscopic image of the grating ridges at the end of a grating acquired by ANT \cite{AppliedNanotools} after etching and b) an optical microscopic image of the mode converter and grating of design A. c) The optical setup used to characterize the designs. Three operational modes can be chosen by inserting/retracting mirrors: (i) optical microscopic imaging, (ii) SWIR microscopic beam imaging and (iii) measuring $M^2$ beam quality.}
\textsuperscript{}
\label{fig:experimental_setup}
\end{figure}

\textbf{Chip fabrication}\\
The chips were fabricated on SOI multi-project wafers by Applied Nanotools (ANT) \cite{AppliedNanotools} and consisted of a Si-handle-layer, a SiO$_2$ buried oxide layer and a Si device layer with thicknesses of 725\,\textmu m, 2\,\textmu m and 220\,nm, respectively. The device layer was patterned (full device layer etch) using 100\,keV electron beam lithography and an SiO$_2$ top oxide cladding with a thickness of 2.2\,\textmu m was deposited on top of the patterned device layer using a plasma-enhanced chemical vapor deposition process. A scanning-electron microscope image of the grating ridges and an optical microscope image of a grating coupler are shown in Figures \ref{fig:experimental_setup}a and \ref{fig:experimental_setup}b, respectively.\\
\textbf{Beam characterization}\\
The experimental setup used to characterize the beam profiles is shown in Figure \ref{fig:experimental_setup}c. The custom microscope setup is built around the Thorlabs Cerna$^{\textregistered}$ microscope platform and uses three different beam paths for visible microscopy (VIS CMOS camera: Thorlabs CS165MU/M), SWIR microscopic beam imaging (SWIR InGaAs camera: NIT WiDy SenS 640) and measuring the $M^2$ beam quality factor. The microscope is equipped with long working distance objectives (Mitutoyo M Plan Apo NIR 5X (NA = 0.14) and Mitutoyo M Plan Apo NIR 10X (NA = 0.26)) with a transmission window of 480\,nm to 1800\,nm and are used together with 200\,mm focal length tube lenses (Thorlabs WFA4100 for VIS and Thorlabs TTL200-S8 for SWIR). In this work only the 10X objective was used since the emission angle of Design A (-9.4°) lay outside of the acceptance cone of the 5X objective, given by $\Theta_\text{max,5X} = \text{arcsin}(\text{NA}) \approx 8.05$°. The chips were mounted on a vacuum chip mount fixed to an optical table. The microscope was placed on a long-travel z-stage (Standa MT195), which was mounted on an xy-stage (Standa 8MTF-200) with step resolutions of 12.5\,\textmu m and 2.5\,\textmu m, respectively. The microscope setup was used to image the beam in the first 80\,mm above the chip. To measure the beam quality factors $M^2$ and beam propagation at larger distances, an additional retractable mirror was placed above the chip to guide the beam parallel to the optical table. For the $M^2$ measurements, the beam was collimated using a plano-convex collimation lens with f$_\text{c}$ = 500\,mm and the x-axis side lobes of the beam were cropped using a spherical 8 mm aperture. The $M^2$ measurements were conducted using the commercial Thorlabs M2 Measurement System M2MS-BP209IR, which consists of two mirrors mounted on a linear stage and a scanning slit beam profiler. The focusing plano-convex lens shipped with the system had a focal length of f$_{M^2}$ = 250\,mm and is supposed to allow the beam profiler to image the beam around its focus by moving the linear stage. However, the divergences of the beams under test were too large, causing the focus to be out of the range of the system. Thus, an additional collimating lens with f$_\text{c}$ = 500\,mm was placed at a distance of 500\,mm from the chip to decrease the divergence of the beam, which placed the focus in the center of the stage's range (see Figure \ref{fig:M2} for reference). The x- and y-axes of the beam profiler were aligned with the x- and y-axes of the beam (see Figure \ref{fig:full_beams} for reference). Subsequently, the linear stage scanned its entire range and allowed the beam profiler to track the intensity profiles along the two axes and calculate the waist diameters along x and y at every position. All $M^2$ measurements were ISO 11146 compliant. For measuring the beam propagation in the far field, the $M^2$ setup including the collimation lens was removed and the SWIR camera was placed in the beam path at distances of 57\,cm to 72\,cm from the chip.\\
%%%%%%%% Figure start %%%%%%%%%
%%%%%%%% Figure end %%%%%%%%%
Designs A and B were located on two different chips and used different edge coupling methods: For design A, the chip was equipped with facet-attached micro lenses (FaMLs) from DreamPhotonics \cite{DreamPhotonics} which facilitate the alignment of the chip and fiber-array. Consequently, a lensed, polarization maintaining fiber array from DreamPhotonics was used for in-coupling. For the case of an SOI-chip fabricated by Applied Nanotools with the additional FaMLs, DreamPhotonics reports an average insertion efficiency of $T = 64.2\,\%$ for TE polarization. For TM polarization DreamPhotonics estimates an additional polarization-dependent loss of  11\,\%, resulting in a total coupling efficiency of  $T = 64.2\,\%\cdot 89\,\%  =57.1\,\%$. Additionally, a power splitter with an experimentally determined splitting ratio of 18.7\,\%/81.3\,\% was placed before the waveguide reaches the evanescent mode converter. Thus the total transmission from fiber array to the grating (assuming T = 100\,\% for the mode converter) is $T = 57.1\,\% \cdot 18.7\,\% = 10.7\,\%$. The power emitted from the fiber array and the emitted power 7\,cm above the chip were measured  to be 2.541\,mW and 83.0\,\textmu W, respectively. Thus, the estimated radiation efficiency of design A is $T_\text{A} =$ 83.0\,\textmu W/ 2,541\,\textmu W / 0.107 $\approx$ 31\,\%. For design B,  nano-tapered edge couplers from the ANT PDK were used with a lensed, focusing PM fiber array from PHIX with a focal mode field diameter (MFD) of 3.6\,\textmu m. ANT reports an average insertion efficiency of T = 51.3\,\% when using a lensed PM fiber with a MFD of 2.5\,\textmu m. Thus, the calculated radiation efficiency for design B represents a lower boundary for the actual efficiency. The power emitted from the fiber array and the power 7\,cm above the grating were measured to be 2.963\,mW and 0.744\,mW, respectively. The resulting lower boundary for the radiation efficiency is therefore $T_\text{B} \geq $ 0.744\,mW/ 2.963\,mW / 0.513 $\approx$ 49\,\%.\\

\section{Conclusion}
We have demonstrated silicon-on-insulator grating couplers that emit collimated Gaussian beams with diameters of several hundred micrometers at $\lambda = 1.55$\,\textmu m. These couplers are designed by tailoring the local coupling strength of a sub-wavelength grating along its length and independently controlling the local emission angle. Beam quality measurements, conducted in accordance with ISO 11146 \cite{ISO11146-1}, yield $M^2 \leq 1.10$ ($\pm$5\,\%). Although large-area grating couplers have previously been reported, their beam quality has, to our knowledge, not been quantified. The present work establishes near-ideal beam quality as an achievable and verifiable property of this coupler class.

The framework can be directly adapted to other target field distributions, such as flat-top beams or higher-order modes, by substituting the desired complex electric field in Equation \ref{eq:E_field}. This adaptability makes meta-grating couplers a practical chip-to-free-space interface for applications sensitive to mode matching, including coupling into high-finesse resonators and forming optical traps. A logical next step is to compensate for residual emission-angle variation using metasurface-based wavefront correction, which would further enhance the achievable mode overlap.\\
\medskip
\textbf{Supporting Information} \par 
Supporting Information is available from the author.

% Acknowledgements
\medskip
\textbf{Acknowledgements} \par %delete if not applicable))
This project has received funding from the European Research Council (ERC) under the European Union’s Horizon Europe research and innovation program (Project Number
[101170022]). This work is partially funded by the Deutsche Forschungsgemeinschaft (DFG, German Research Foundation) under Germany’s Excellence Strategy – EXC-2123 Quantum-Frontiers – 390837967. M.S. acknowledges support from the Max Planck School of Photonics.

\newpage
\textbf{Supplementary Information}

\begin{figure}[h]
\centering  
\includegraphics[width=0.8\linewidth]{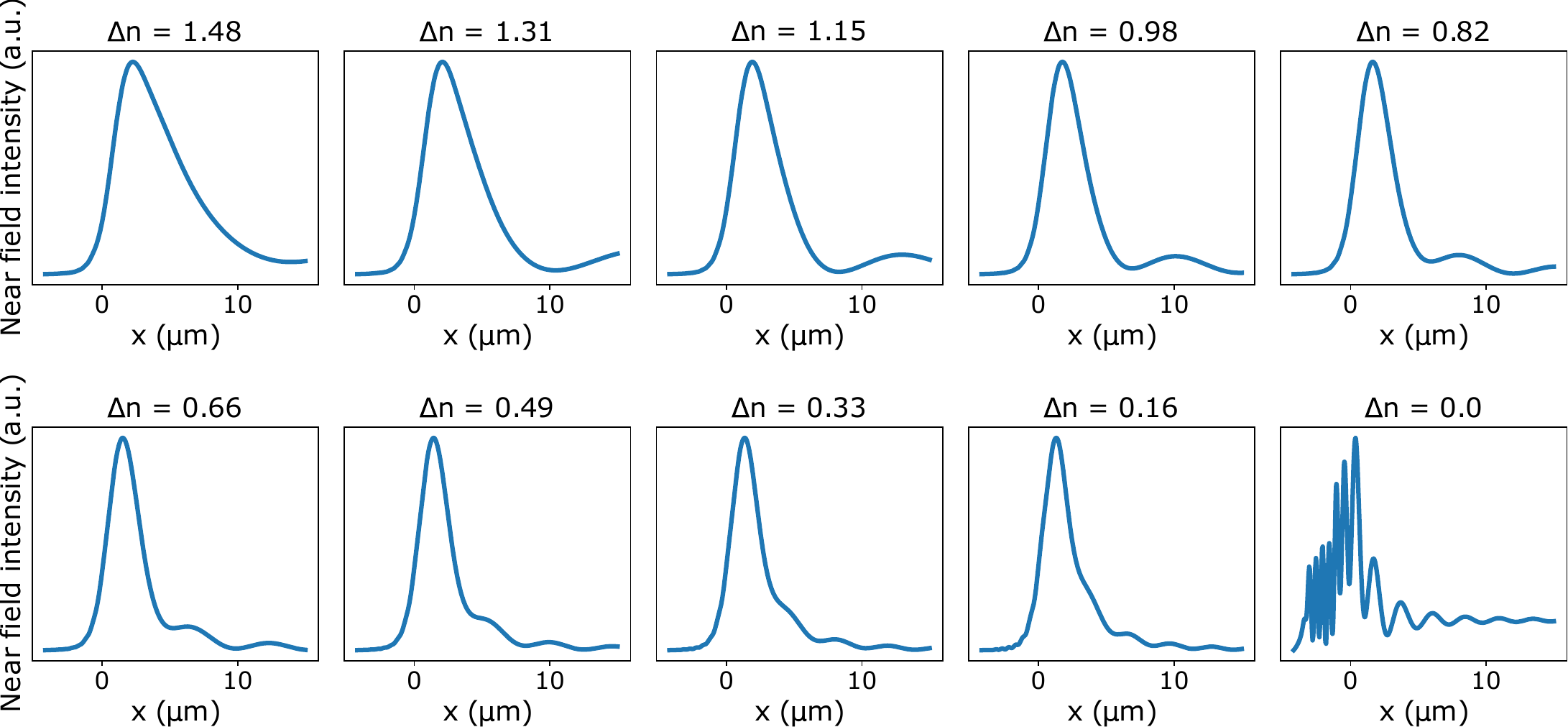}
\caption{2D-FDTD-simulations of normalized near field intensities above a waveguide as a function of $\Delta n = n_1-n_2$, which abruptly changes its refractive index from silicon ($n_1$\,=\,$3.476$) to a lower index of $n_2$ at x = 0. The waveguide's height is 220 nm and the top and bottom oxide layers consist of SiO$_2$ with a refractive index of 1.444 at a wavelength of $\lambda = 1.55$\,\textmu m. These simulations show the origin of the emission peak of the grating designs shown in this contribution. The polarization was set to TM.}
\textsuperscript{}
\label{fig:waveguide_transition_fields}
\end{figure}

\begin{figure}[h]
\centering  
\includegraphics[width=0.3\linewidth]{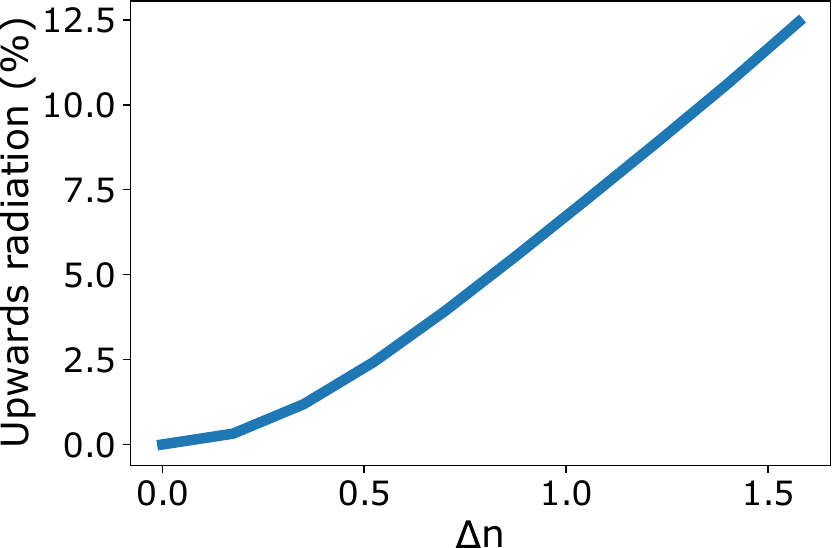}
\caption{The upwards (+z) emitted radiation resulting from the abrupt waveguide transitions from Figure \ref{fig:waveguide_transition_fields} as a function of $\Delta n$.}
\textsuperscript{}
\label{fig:waveguide_transition_Tvals}
\end{figure}

Figures \ref{fig:waveguide_transition_fields} and \ref{fig:waveguide_transition_Tvals} show the origin of the near field emission peak observed in both the simulations and experimental measurements of the meta-grating designs. Due to the abrupt refractive index change at the beginning of the grating non-directional emission can be observed. The amount of power that gets radiated depends on the difference in refractive index (Figure \ref{fig:waveguide_transition_Tvals}). For the designs discussed in this contribution, the refractive index of the slab waveguide is that of silicon, i.e. $n_1 = 3.476$. The effective refractive index of the meta-grating can be estimated since the average duty cycle of each cell $DC_\text{mean}$ remains approximately constant along the grating. The effective refractive index for TM-polarized light can then be estimated by \cite{Cheben2018}: 
\begin{equation}
    n_\text{eff} \approx \frac{n_1\cdot n_2}{\sqrt{DC\cdot n_2^2 + (1-DC)\cdot n_1^2}}
\end{equation}
For designs A and B, the average duty cycles of $DC_\text{mean}^\text{A} = 0.5$ and  $DC_\text{mean}^\text{B} = 0.625$ yield effective indices of $n_\text{eff}^\text{A} = 1.886$ and $n_\text{eff}^\text{B} = 2.079$, corresponding to refractive index jumps of $\Delta n^\text{A} = 3.476 - 1.886 = 1.590$ and $\Delta n^\text{B} = 3.476- 2.079 = 1.397$. According to Figure \ref{fig:waveguide_transition_Tvals}, these jumps cause parasitic emissions of 13\,\% and 10\,\% for designs A and B, respectively.

%%%%%%%% Figure start %%%%%%%%%
%%%%%%%% Figure end %%%%%%%%%

%%%%%%%% Figure start %%%%%%%%%
%%%%%%%% Figure end %%%%%%%%%

% References
\medskip

\bibliographystyle{unsrt}
\bibliography{literature.bib}

\end{document}